\documentclass[aps,prl,twocolumn,showpacs,amsmath,amssymb,
groupedaddress]{revtex4}

\usepackage{graphicx}

\begin{document}


\title{Realizability of stationary spherically symmetric transonic
accretion}

\author{Arnab K. Ray}
\author{J. K. Bhattacharjee}
\affiliation{Department of Theoretical Physics \\
Indian Association for the Cultivation of Science \\
Jadavpur, Calcutta (Kolkata) 700032, INDIA}

\date{\today}  

\begin{abstract}
The spherically symmetric stationary transonic (Bondi) flow is 
considered a classic example of an accretion flow. This flow, 
however, is along a separatrix, which is usually not physically 
realizable. We demonstrate, using a pedagogical example, that it 
is the dynamics which selects the transonic flow. 
\end{abstract}


\maketitle

In astrophysics, the importance of accretion processes can
hardly be overstated, especially in the context of the study
of compact astrophysical objects and active galactic nuclei.
Of such processes, a very important paradigm is that of the steady
spherically symmetric flow, in which the motion of the accreting
matter is steady and spherically symmetric, obeying the boundary 
condition that the bulk velocity is to fall off to zero at infinity,
while the density asymptotically approaches a fixed value. Since this 
is the simplest situation in the class of accretion flows, it is the 
starting point in all relevant texts \cite{skc90,fkr92}. Studied 
extensively, among others, by Bondi \cite{bon52}, about fifty years 
ago, it also gives one of the clearest examples of a transonic flow 
(with which, in spherical symmetry, Bondi's name is associated),
i.e. a flow in which the velocity is subsonic far away from
the star and becomes supersonic as the surface of the star is
approached. It is this simplicity and easy comprehensibility
of the model, that overshadows the fact that it is not encountered 
most often in practice. Accordingly, the spherically symmetric flow 
has been of continuing interest \cite{pso80, td92, csw96, tor99}.
We first argue that the transonic (Bondi) solution of the steady 
spherically symmetric flow, would not be physically realizable. 
This proposition is in contradiction to the position of Bondi
himself, that ``the case physically most likely to occur is 
that with the maximum rate of accretion" \cite{bon52}, which, 
in spherically symmetric accretion, is a case that 
is readily identified as the transonic solution \cite{pso80}. 
We then argue that it is the dynamics which actually selects
the transonic flow from among all possible trajectories. 

\begin{figure}[b]
\begin{center}
\includegraphics[scale=0.9, angle=0]{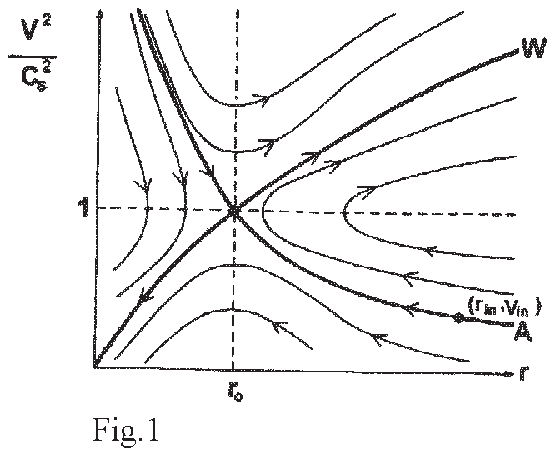}
\caption{\label{f.1} \small{Phase trajectories for spherically symmetric
accretion onto a star. The bold solid lines, {\bf A} and {\bf W}, 
represent ``accretion" and ``wind" respectively. The fixed point is 
at $r=r_0$ and  $v^2/{c_{\mathrm{s}}}^2=1$. 
Linear stability analysis 
indicates that the fixed point of the flow is a saddle point. The 
direction of the arrows along the line {\bf A}, demonstrates that the 
transonic flow is not physically realizable.}}
\end{center}
\end{figure}

The typical solutions for the spherically symmetric flow in the
velocity-coordinate space are shown in Fig.\ref{f.1} (to be read without 
the arrows). The two dark solid curves labelled {\bf A} and {\bf W} refer 
to the accretion flow and the wind flow respectively. The intersection
point is an equilibrium point. The problem with Fig.\ref{f.1} is that 
when it is read without the arrows it is slightly misleading. It does
not show along what route an integration of $dv/dr$ would proceed if we 
start with an initial condition $v=v_{\mathrm{in}}$
at $r=r_{\mathrm{in}}$ far away from the star. For a physically 
realizable flow, an initial condition infinitesimally close to a point 
on the accretion curve {\bf A}, would trace out a curve infinitesimally
close to {\bf A} and in the limit would correctly reproduce {\bf A},
evolving along it and passing through the equilibrium point (sonic
point) as we integrate $dv/dr$, obtained from Euler's
equation. We will show that the arrows on the integration route
are as shown in Fig.\ref{f.1}. The direction of the arrows indicates
that the spherically symmetric transonic accretion flow
should not be physically realizable.

Is this a result confined to the spherically symmetric flow? We have 
verified, as we shall show briefly, that this holds also for the 
axisymmetric rotating accretion flow for a thin disc \cite{ny94} which 
is a situation of practical interest. The technique for assigning the 
arrows remains identical to that of the spherically 
symmetric case \cite{rb02}.

\begin{figure}
\begin{center}
\includegraphics[scale=0.8, angle=0]{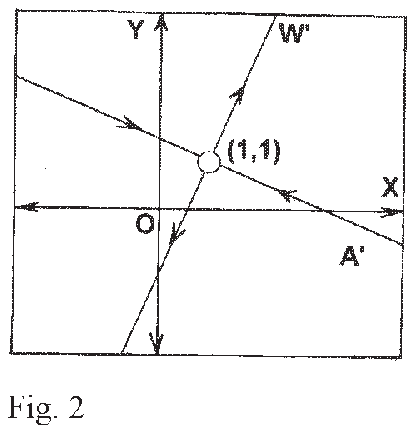}
\caption{\label{f.2} \small{Integration of Eq.(\ref{toystat}) 
gives a pair of straight
lines, with the integration constant fixed by the intersection point $(1,1)$. 
In the figure the lines are marked ${\bf A}^{\prime}$ and ${\bf W}^{\prime}$.
Parametrizing Eq.(\ref{toystat}) as Eq.(\ref{toypara}), we get the concept 
of arrows. Linear 
stability analysis indicates that for ${\bf A}^{\prime}$ and 
${\bf W}^{\prime}$, the intersection point $(1,1)$ is actually a saddle 
point, for which the arrows are as shown above.}}
\end{center}
\end{figure}

We wish to make our point with a fairly straightforward example. We
consider the differential equation

\begin{equation}
\label{toystat}
\frac{dy}{dx} = \frac{Y(x,y)}{X(x,y)} = \frac{x+y-2}{y-x}
\end{equation}
whose integral can be written down as 

\begin{equation}
\label{integtoy}
{x^2} - {y^2} - 4x + 2xy = - C
\end{equation}
where $C$ is a constant. If we want that particular solution which 
passes through the point where $Y(x,y)=X(x,y)=0$, namely $x=y=1$,
then $C=2$. The curve $x^2-y^2-4x+2xy=-2$ factorizes into a 
pair of straight lines : $y-x(\sqrt{2}+1)+\sqrt{2}=0$ and
$y+x(\sqrt{2}-1)-\sqrt{2}=0$. This pair is shown in Fig.\ref{f.2} 
(to be read again without the arrows) as the
lines marked ${\bf A}^{\prime}$ and ${\bf W}^{\prime}$.

We want to explore the process of drawing the line ${\bf A}^{\prime}$
from a given initial condition. On line ${\bf A}^{\prime}$, 
$y=0$ at $x=2+\sqrt{2}$. Let us begin with the initial condition $y=0$ 
at $x=2+\sqrt{2}-\epsilon$, where $0< \epsilon \ll 1$. The given initial 
condition fixes the 
constant $C$ as $C=2[1+\sqrt{2}\epsilon-{\epsilon}^2/2]$. 
Using this value of $C$, we can plot the curve given by 
Eq.(\ref{integtoy}).  
For a given $x$, a value of $y$ is given by the relevant root of the 
quadratic equation thus obtained. The two roots are  

\begin{equation}
\label{toyroot}
y = x \pm \sqrt{2} \bigg[ (x-1)^2 + \sqrt{2} \epsilon - 
\frac{{\epsilon}^2}{2} \bigg]^{1/2}
\end{equation}
from which it is clear that to satisfy $y=0$ at $x=2+\sqrt{2}-\epsilon$, 
the negative sign has to be chosen to let us have  

\begin{equation}
\label{rootneg}
y = x - \sqrt{2} \bigg[ (x-1)^2 + \sqrt{2} \epsilon -
\frac{{\epsilon}^2}{2} \bigg]^{1/2}
\end{equation}
At $x=0$, we then get 
$y = - \sqrt{2}(1+\sqrt{2} \epsilon - {\epsilon}^2/2)^{1/2}$, which is
very different from $y=\sqrt{2}$, that one gets on line 
${\bf A}^{\prime}$. In the limit of $\epsilon \longrightarrow 0$, one 
generates a part of ${\bf A}^{\prime}$ $(x \geqslant 1)$ and a part of 
${\bf W}^{\prime}$ $(x \leqslant 1)$, instead of the entire line
${\bf A}^{\prime}$. Another way of stating this is the sensitivity to 
initial conditions in the drawing of ${\bf A}^{\prime}$. 
If we make an error of an infinitesimal amount
$\epsilon$ in prescribing the initial condition on ${\bf A}^{\prime}$,
i.e. if we prescribe $y=0$ at $x=2+\sqrt{2}-\epsilon$ instead of $y=0$
at $x=2+\sqrt{2}$, then the ``error" made at $x=0$ relative to 
${\bf A}^{\prime}$ is $2\sqrt{2}$ which is ${\mathcal{O}}(1)$. An 
infinitesimal separation at one point leads to a finite separation at 
a point a short distance away. This is what we mean by saying that  
line ${\bf A}^{\prime}$ (and similarly ${\bf W}^{\prime}$) should not 
be physically realized.

The clearest and most direct understanding of the difficulty is
achieved by writing Eq.(\ref{toystat}) as a set of two parametrized 
differential equations

\begin{align}
\label{toypara}
\frac {dy}{d \tau} &= x + y -2 \nonumber \\
\frac {dx}{d \tau} &= y - x
\end{align}
where $\tau$ is some convenient parameter. The fixed point
of this dynamical system is $(1,1)$, namely the point where $Y(x,y)$
and $X(x,y)$ are simultaneously zero --- the point through which both
${\bf A}^{\prime}$ and ${\bf W}^{\prime}$ pass. Linear stability 
analysis of this fixed point in $\tau$ space 
shows that it is a saddle. The critical solutions 
in this $y$ --- $x$ space can now be drawn with arrows and the 
result is as shown in Fig.\ref{f.2}. The distribution of the arrows 
(characteristic of a saddle \cite{js77}) implies ${\bf A}^{\prime}$ 
and ${\bf W}^{\prime}$ cannot be physically realized. 

For the spherically symmetric flow, the relevant variables are the 
radial velocity $v$ and the local density $\rho$. Ignoring viscosity, 
we write down Euler's equation for $v$ as

\begin{equation} 
\label{euler}
\frac{\partial v}{\partial t} + v\frac{\partial v}{\partial r} =
- \frac{1}{\rho}\frac{\partial P}{\partial r}
- \frac{\partial V}{\partial r}
\end{equation}
where $P$ is the local pressure and $V \equiv V(r)$ is the potential 
due to a gravitational attractor of mass $M$, i.e. $V(r)=-GM/r$.
The pressure is related to the local density through the equation
of state $P=K{\rho}^{\gamma}$ where $K$ is a constant, and $\gamma$ 
is the polytropic exponent with an admissible range given by 
$1< \gamma < 5/3$. The local density evolves according to 
the equation of continuity,

\begin{equation}
\label{con}
\frac{\partial \rho}{\partial t} + \frac{1}{r^2}\frac{\partial}
{\partial r}\big(\rho vr^2 \big) = 0
\end{equation}
The stationary solution implies ${\partial v}/{\partial t} =
{\partial \rho}/{\partial t} = 0$. The local sound speed is given by
${c_{\mathrm{s}}}^2 = {\partial P}/{\partial \rho} 
= \gamma K \rho^{\gamma - 1}$ 
and in the stationary situation, we can use Eq.(\ref{con}) to write 
Eq.(\ref{euler}) as 

\begin{equation}
\label{dvdr}
\frac{d(v^2)}{dr} = \frac{2 v^2}{r}\frac{\big(2{c_{\mathrm{s}}}^2 - 
GM/r \big)}{v^2 - {c_{\mathrm{s}}}^2}
\end{equation}
whose integration with different initial conditions is supposed to 
generate the curves in Fig.\ref{f.1}. To assign arrows we 
write Eq.(\ref{dvdr}) in a parametrized form

\begin{align}
\label{paradvdr}
\frac{d(v^2)}{d \tau} &= 2v^2 \bigg( 2{c_{\mathrm{s}}}^2 - 
\frac{GM}{r} \bigg) \nonumber \\
\frac{dr}{d \tau} &= r \big( v^2 - {c_{\mathrm{s}}}^2 \big)
\end{align}
to find that the critical point is at $r=r_0$ and
$v=v_0$, such that 
${v_0}^2={c_{\mathrm{s0}}}^2$ and ${c_{\mathrm{s0}}}^2=
GM/2r_0$. It is immediately clear that the critical point is 
the so called sonic point, and it can be fixed in terms of the constants
of the system with the help of suitable boundary conditions \cite{skc90}. 
We now carry out a linear stability analysis around the fixed point by 
writing $v^2={v_0}^2(1+{\delta}_1)$ and $r=r_0 (1+{\delta}_2)$.
Linearizing in ${\delta}_1$ and ${\delta}_2$, followed by some 
straightforward algebra then yields 

\begin{align}
\label{linper}
\frac{d {\delta}_1}{d \tau} &= 2{c_{\mathrm{s0}}}^2 
\Big [-(\gamma -1){\delta}_1 
+ (6-4 \gamma ){\delta}_2 \Big ] \nonumber \\
\frac{d {\delta}_2}{d \tau} &= {c_{\mathrm{s0}}}^2
\bigg [ \frac{\gamma +1}{2}{\delta}_1 + 2(\gamma -1){\delta}_2 \bigg ]
\end{align}

The solutions for ${\delta}_{1,2}$ are of the form $e^{\lambda \tau}$,
where $\lambda$ is to be found from the roots of

\begin{equation}
\label{deter}
\begin{vmatrix} 
\lambda + 2(\gamma-1){c_{\mathrm{s0}}}^2 & 
2(4\gamma-6) {c_{\mathrm{s0}}}^2 \\ 
-[(\gamma +1)/2]{c_{\mathrm{s0}}}^2 & 
\lambda -2(\gamma -1) {c_{\mathrm{s0}}}^2
\end{vmatrix} = 0
\end{equation}

The eigenvalues $\lambda$ are given by $\lambda=\pm {c_{\mathrm{s0}}}^2 
\sqrt{2(5-3\gamma)}$. For $\gamma < 5/3$, the fixed point is a saddle.
Since the physical situations are restricted to this range, we
find that the fixed point is always a saddle and hence the
distribution of arrows will be as in Fig.\ref{f.1}.

For the more realistic axisymmetric case (as in a thin accretion disc), 
using $R$ as the radial distance in the plane, the static equations of the 
flow for an angular velocity $\Omega$ and for the disc thickness $H$, are

\begin{itemize}

\item[]
The Equation of Continuity :

\begin{equation}
\label{discon}
\rho RHv= \mathrm{constant}
\end{equation}

\item[]
The Momentum Balance Equation : 

\begin{equation}
\label{dismom}
\frac{1}{2} \frac{d(v^2)}{dR}=R{\Omega}^2  - R{{\Omega}_{\mathrm{K}}}^2  
- \frac{1}{\rho} \frac{dP}{dR}
\end{equation}

\item[]
The Angular Momentum Balance Equation :

\begin{equation}
\label{disang}
v \frac{d}{dR} \big(\Omega R^2 \big)= \frac{1}{\rho RH} \frac{d}{dR} 
\bigg ( \frac{\alpha \rho {c_{\mathrm{s}}}^2 R^3 H}{{\Omega}_{\mathrm{K}}} 
\frac{d \Omega} {dR} \bigg ) 
\end{equation}

\end{itemize}
where ${{\Omega}_{\mathrm{K}}}^2=GM/R^3$ and $\alpha$ is the effective 
viscosity of Shakura and Sunyaev \cite{ny94}. An integral of motion follows 
from Eq.(\ref{dismom}) and we can obtain an equation for $d(v^2)/dR$
akin to Eq.(\ref{dvdr}), which we can write as the dynamical system

\begin{align}
\label{dispara}
\frac{d(v^2)}{d \tau} &= 2v^2 \bigg[\frac{5{c_{\mathrm{s}}}^2}{1+ \gamma}
- R^2 \big({{\Omega}_{\mathrm{K}}}^2-{\Omega}^2 \big) \bigg] \nonumber \\
\frac{dR}{d \tau} &= R \bigg [v^2-\frac{2{c_{\mathrm{s}}}^2}
{1+ \gamma} \bigg ]
\end{align}
from which, taking the inviscid limit (i.e. $\alpha = 0$),
an analysis identical to the one following 
Eq.(\ref{paradvdr}) leads to the 
eigenvalues $\lambda$ of the stability matrix. Of these one can 
easily be shown to be appropriate for a saddle \cite{rb02}. 

We now focus on the fact that the real physical problem is dynamic 
in nature as exhibited in Eqs.(\ref{euler}) and (\ref{con}), even 
as we have tacitly assumed that we can study the problem directly 
in its stationary limit. As it turns out there are an infinite 
number of stationary solutions and if the static limit is taken 
directly, the Bondi solution becomes very sensitive to the process 
of integration to determine the stationary
trajectory. Several studies in the linear stability analysis
of the stationary flows that have been carried out in the past
have not really helped clarify the situation as it was finally
established that among the stationary solutions, the transonic
Bondi solution as well as all the subsonic solutions are stable
in the linear stability sense \cite{pso80, gar79}. Any selection 
mechanism then must be of a non-perturbative nature.

Accordingly, we return to our pedagogic example of Eq.(\ref{toystat}) 
but now consider $y$ as a field $y(x,t)$ with the evolution

\begin{equation}
\label{toydyn}
\frac{\partial y}{\partial t} +(y-x) \frac{\partial y}
{\partial x} = y+x-2
\end{equation}

The stationary solutions $y(x)$ satisfy Eq.(\ref{toystat}) and 
Fig.\ref{f.2} shows that their separatrices are given by 
$y-x \big(\sqrt{2}+1 \big)+\sqrt{2}=0$ and
$y+x \big(\sqrt{2}-1 \big)-\sqrt{2}=0$. It should be easy to see 
that a linear stability analysis in $t$ around the family of stationary 
solutions $y(x)$ would show an infinite number of them to be stable,
and therefore each one of them could be a perfectly valid physical
solution. We will now show that the non-perturbative dynamics actually 
selects the separatrices. 

The general solution of Eq.(\ref{toydyn}) can be obtained by the method 
of characteristics \cite{deb97}. The two independent characteristic 
curves of Eq.(\ref{toydyn}) are obtained from 

\begin{equation}
\label{char}
\frac{dt}{1}=\frac{dx}{y-x}=\frac{dy}{y+x-2}
\end{equation}

They are

\begin{align}
\label{charsol}
y^2 -2xy- x^2 +4x &= C \nonumber \\
\bigg [ x -1 \pm \frac{1}{\sqrt{2}} (y-x) \bigg ]
e^{\mp \sqrt{2}t} &= \tilde{C}
\end{align}
in which the latter is obtained by integrating the ${dx}/{dt}$ 
integral with the help of the first solution. With the choice of the 
upper sign now (since we should want the evolution to proceed through
a positive range of $t$), the solutions of Eq.(\ref{toydyn}) can be 
written as 

\begin{equation}
\label{toysol}
y^2 -2xy- x^2 +4x = f\bigg(\bigg [ x-1 + \frac{1}{\sqrt{2}}(y-x)\bigg]
e^{- \sqrt{2}t} \bigg )
\end{equation}
where $f$ is an arbitrary function, whose form is to be determined
from initial conditions. In this case we choose the initial 
condition that $y(x)=0$ at $t=0$ for all $x$. This leads to 
$f \big( x \big[1- 1/{\sqrt{2}}\big] -1 \big) = - x^2 + 4x$ to give a form 
for $f$ as $f(z)= -2 \big(2 \sqrt{2} +3 \big)z^2 
- 4 \big( \sqrt{2}+1 \big) z + 2$. The solution to Eq.(\ref{toydyn}) 
is then given as

\begin{align}
\label{toyfinsol}
\Big[ y-x \big(\sqrt{2}+1 \big)+\sqrt{2} \Big] 
\Big[ y+x \big(\sqrt{2}-1 \big)-\sqrt{2} \Big] \nonumber \\
= - 2 \big(2+\sqrt{2} \big) \Big[ x \big(\sqrt{2}-1 \big)
+ y - \sqrt{2} \Big]& e^{- \sqrt{2}t} \nonumber \\ 
-\big(3 +2 \sqrt{2} \big) \Big[ x \big(\sqrt{2}-1 \big)
+ y - \sqrt{2} \Big]^2& e^{-2 \sqrt{2}t}
\end{align}

Clearly for $t \longrightarrow \infty$, the right hand side 
in Eq. (\ref{toyfinsol}) tends to
zero and we approach one of the two separatrices shown in Fig.\ref{f.2}. 
Of the two possible separatrices, the one which
will be relevant will be determined by some other requirement. 
For the astrophysical flow the two separatrices are the Bondi accretion
flow and the transonic wind solution. One chooses the proper sign of the 
velocity to get the flow in which one is interested.

The mechanism for the selection of the asymptotes in Fig.\ref{f.1} as 
the favoured trajectories is identical. This can be appreciated from a
look at Eq.(\ref{euler}). In the ``pressure-free" approximation, we 
have a set of stationary solutions $v^2 - 2GM/r = c^2$. 
Which of these would be selected by the dynamics? If we start from $v=0$ 
at $t=0$ for all $r$, an identical reasoning to the one given above
shows that it is the path with $c=0$ which is selected. 
The solution of the differential equation 

\begin{equation}
\label{presfree}
\frac{\partial v}{\partial t} + v \frac{\partial v}{\partial r} =
- \frac{GM}{r^2}
\end{equation}
by the method of characteristics yields 

\begin{equation}
\label{presfreedyn}
\frac{v^2}{2} - \frac{GM}{r} = - \frac{GM}{r} \bigg(\frac{v}{c} + 1
\bigg)^{-2} \exp \bigg( \frac{2rv}{cr_{\mathrm{c}}} - 
\frac{2ct}{r_{\mathrm{c}}} \bigg)
\end{equation}
in which $r_{\mathrm{c}} = 2GM/c^2$. Obviously 
for $t \longrightarrow \infty$ we will get the static solution 
$v^2=2GM/r$. This is also the choice according to the energy criterion. 
Corresponding to the initial condition given, the lowest possible total 
energy is $E={v^2}/2- GM/r =0$ and the dynamics selects this particular 
stationary trajectory. This selection mechanism is entirely non-perturbative 
in character and provides the mathematical justification for Bondi's 
assertion that the energy criterion should select the stationary 
flow \cite{bon52}. With the pressure term included, this argument leads to 
the selection of the transonic path i.e. gravity always wins over pressure
at small radial distances.

\begin{acknowledgments}
This research has made use of NASA's Astrophysics Data System.
The authors would like to thank an anonymous referee for his/her
useful comments. A conversation with S. Sengupta is also 
gratefully acknowledged. One of the authors (AKR) would also 
like to acknowledge the financial assistance given to him by
the Council of Scientific and Industrial Research, Government
of India, and the help of A. Bhattacharyay, Dr. T.K. Das and D. Sanyal.
\end{acknowledgments}

\bibliography{arpre}

\end{document}